\documentclass[twocolumn]{revtex4-1}

\usepackage{latexsym}
\usepackage{amsmath}
\usepackage{amsfonts}
\usepackage{amssymb}
\usepackage{amsbsy}
\usepackage{bm}
\usepackage{color}
\usepackage{graphicx}
\usepackage{longtable}
\usepackage{setspace}
\usepackage{color}
\usepackage{hyperref}
\usepackage{soul}

\begin{document}

\begin{titlepage}
\title{Morphological transitions of active Brownian particle aggregates on porous walls}
\author{Suchismita Das, Raghunath Chelakkot}
\email{raghu@phy.iitb.ac.in}
\affiliation{
 Department of Physics, Indian Institute of Technology Bombay, Mumbai - 400076, India.\\ 
}

\begin{abstract}
Motility-induced wall aggregation of Active Brownian Particles (ABPs) is a well-studied phenomenon. Here, we study the aggregation of ABPs on porous walls, which allows the particles to penetrate through at large motility. We show that the active aggregates undergo a morphological transition from a connected dense-phase to disconnected droplets with an increase in wall porosity and the particle self-motility, similar to wetting-dewetting transitions in equilibrium fluids. We show that both morphologically distinct states are stable, and independent of initial conditions at least in some parameter regions. Our analysis reveals that changes in wall porosity affect the intrinsic properties of the aggregates and changes the effective wall-aggregate interfacial tension, consistent with the appearance of the morphological transition. Accordingly, a close analysis of the density, as well as orientational distribution, indicates that the underlying reason for such morphological transitions is not necessarily specific to the systems with porous walls, and it can be possible to observe in a larger class of confined, active systems by tuning the properties of confining walls.
 \end{abstract}
\maketitle
\end{titlepage}

Active matter is a prominent class of non-equilibrium systems where microscopic autonomous motion leads to distinct types of collective ordering~\cite{Marchetti2013, Ramaswamy2010,Vicsek2012}. Such non-equilibrium ordering is ubiquitous in a range of biological~\cite{Ballerini2008,Katz2011,Dombrowski2004,Sanchez2012,Patteson2018,Dell'Arciprete2018,Wioland_2016,Wioland2013,DeCamp2015,Kawaguchi2017}, as well as synthetic systems~\cite{Therkauff2012, Palacci2013, Volpe2011, Buttinoni2013, Bricard2013, Thutupalli2011, Bechinger2016, vanderLinden2019, Nishiguchi2015, Yan2016}. A particular kind of non-equilibrium ordering observed in active matter is the Motility induced phase separation (MIPS), where active elements segregate into a dense and to a dilute phase without any cohesive interactions~\cite{Redner2013,Fily2012,Cugliandolo2017,Farage2015,Klamser2018,Liu2019,herminghaus2017phase, Cates2015, caprini2020spontaneous}. Despite being an intrinsically non-equilibrium process, MIPS shows striking similarities to equilibrium liquid-gas phase coexistence. Due to this property, several concepts of equilibrium statistical physics, such as pressure~\cite{Takatori2014,Winkler2015,Fily_2017,Solon2015,Nikola2016,Marconi2016,Patch2018,Speck2016,Patch2017,Patch_curvature,JamaliNaji2018_anomalousdropletripening,das2019local}, surface tension~\cite{Bialke2015,paliwal2017non,prymidis2016vapour,Solon_2018,Junco_JCP2019_surfacetension,cagnetta2020kinetic,zakine2019surface,omar_PRE2020}, and chemical potential~\cite{takatori2015towards,paliwal2018chemical,Tjhung2018,guioth_jcp2019} have been applied to describe MIPS. In the presence of non-adhesive walls, a formation of dense-phase of active particles is observed on the wall surface as the wall reduces particle motility~\cite{lee2017interface,yang2014aggregation,sepulveda2017wetting,SepulvedaSoto_PRE2018_activewetting,ni2015tunable,Kne_evi__2020}.

\begin{figure}
	\includegraphics[width=0.98\columnwidth]{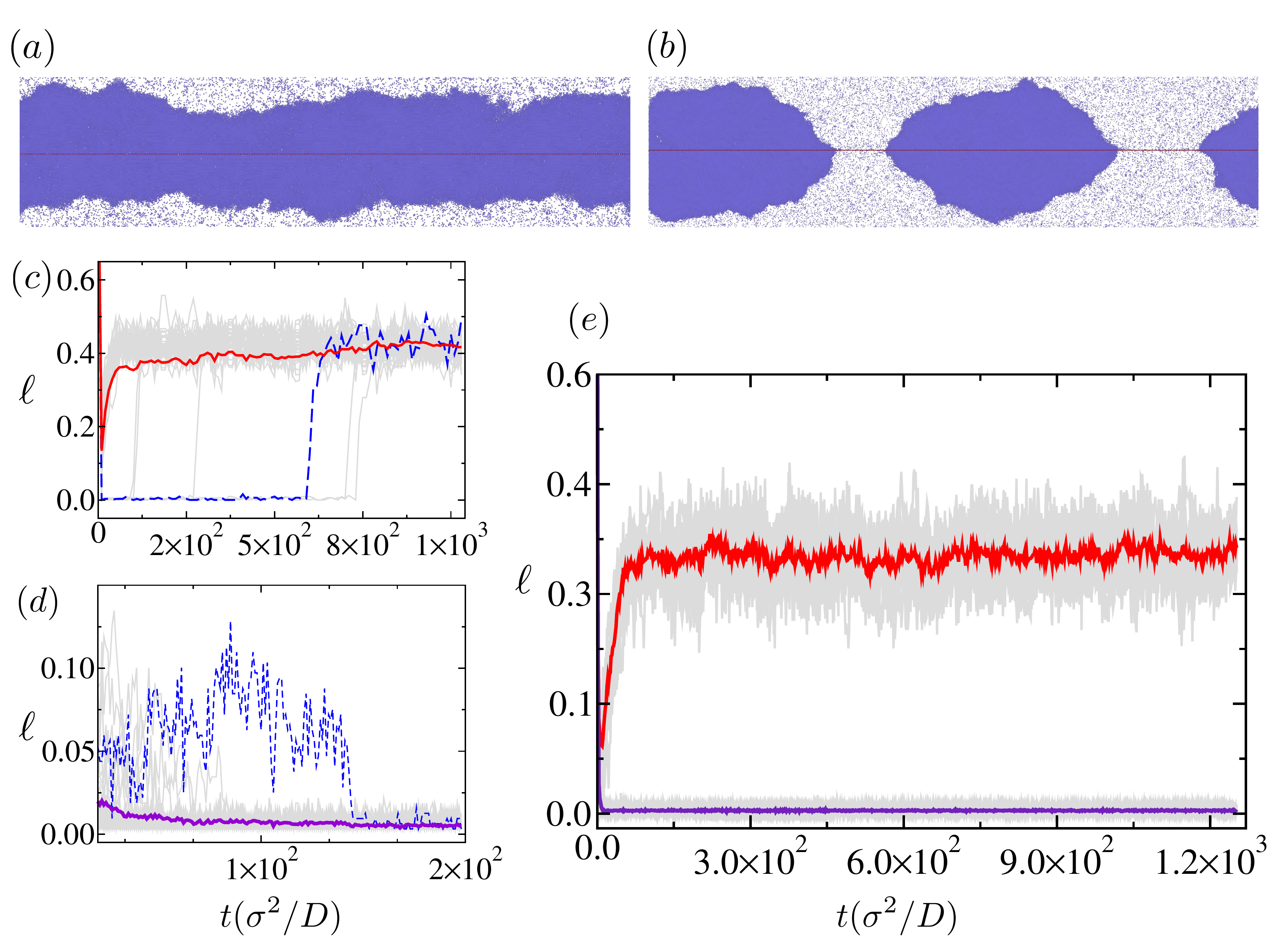}
	\caption{Examples of a connected dense-phase (CDP) (a) ($Pe=120$, $\delta = 1.59$) (\cite{SI}-Movie\S1) and (b) disconnected droplets ($Pe=240$, $\delta = 1.85$) (\cite{SI}-Movie\S2) with $N= 249500$. Time evolution of the vacancy fraction (fraction of cluster free region) on the wall ($\ell$) at $N=39800$ for (c) $\delta = 1.85 $, $Pe = 240$, all the individual trajectories eventually converge to droplets with $\tilde \ell \simeq 0.4$ with late-time transitions (\cite{SI}-Movie\S3). (d) $\delta = 1.69 $, $Pe = 120$, all the trajectories form continuous clusters ($\ell =0$) with a few late-time transitions (\cite{SI}-Movie\S4), and (e) $\delta = 1.85$, $Pe = 200$, showing robust bistable behaviour with $26$ out of $40$  simulations converging into CDP. The grey lines represent individual simulations, red lines are the average $\ell$ over discontinuous droplets, and violet lines represent the average  $\ell$ over the runs that form CDP. The blue line in (c) and (d) denotes a typical transitioning trajectory. 
	}
	\label{fig:1}
\end{figure}

Active particle aggregation on solid surfaces has a special interest in the context of understanding microbial dynamics near walls~\cite{Lauga2006, Elgeti2013}, and the formation of biofilms ~\cite{hall2004bacterial,keller2019, Hartmann2019, Liu2019}. {In the case of simple excluded volume interactions between solid surfaces and particles, the aggregates spread almost uniformly on the wall~\cite{ni2015tunable, lee2017interface}. Recent studies have shown qualitative similarities to capillary condensation in equilibrium fluids~\cite{wysocki2020capillary,Kne_evi__2020}}. It has also been shown that the reorientation rate of active particles changes the wall aggregation properties~\cite{sepulveda2017wetting, SepulvedaSoto_PRE2018_activewetting}.
 Recently, there is a growing interest in studying the targeted assembly of active particles, either by varying the interactions or the environment. One of the major objectives of such studies is to find an optimum arrangement of excluded regions in space, to achieve aggregates of desired shape and size. So far, the focus of such studies has been to optimize the geometry, distribution, and the mutual orientation angles between smooth, impenetrable obstacles to facilitate particle trapping~\cite{Kaiser2012, Kumar2019, Magiera_PRE2015_trapping, Wu_2018_transport_barrier, Reichhardt_2017_activetrapping}. However, the possibility of tuning the wall-particle interaction to control the self-assembly has not been explored much. One possible way to change the physical interaction between the particles and the wall is to introduce wall penetrability. Recent studies showing particle trapping by penetrable membranes also indicate possible biomedical applications of such studies ~\cite{Daddi_Moussa_Ider_2019,daddi2019membrane}.

To study the effect of wall penetrability in aggregate morphology, we simulate the dynamics of Active Brownian particles (ABPs) in the presence of a rigid and porous wall. We show that the dense-phase on the wall surface displays two morphologically distinct states, a connected dense-phase (CDP), where the dense-phase spreads entirely on the wall (Fig~\ref{fig:1}(a)) (Movie\S1 \cite{SI})), and disconnected droplets (Fig.~\ref{fig:1}(b)) (Movie\S2 \cite{SI})) where complete spreading is prevented by the formation of multiple finite-sized aggregates with a macroscopically curved interface. The CDP state is similar in structure to the cluster formation observed on the surface of perfectly impenetrable, smooth walls~\cite{ni2015tunable, Kne_evi__2020}.
We observe that the transition from one morphological state to the other happens when either the pore size or the particle motility is varied. Also, for a large enough ABP motility, changing the wall porosity alone leads to a morphological transition. 
To analyze this phenomenon, we calculate the mechanical tension at the solid-liquid interface, following the method used in equilibrium interfaces ~\cite{Kirkwood1949, nijmeijer1990wetting}, recently extended for active systems~\cite{Bialke2015,paliwal2017non}. We show that the preference to form disconnected droplets over a connected dense-phase concurs with an increase in solid-liquid interfacial tension, as the wall changes the internal properties of the aggregates.

We study a two-dimensional system of $N$ disk-like particles in a square simulation box (length $L$) with periodic boundary conditions. The particles interact via short-ranged repulsive, Weeks-Chandler-Anderson (WCA)~\cite{Weeks1971} potential.
The porous wall is modelled by linearly arranging static `wall' particles at a uniform separation $d$ at the center of the simulation box along the $x$ direction (at $y$ = L/2). The particle-wall interaction  $U^{\text{pw}}$ is also modeled by WCA potential. 
The dynamics of the $i^{\text{th}}$ particle's position $\textbf{r}_i$ is governed by,
\begin{equation}
{\dot{\bf{r}}}_i  = -\mu{\bm\nabla}_i\left(U^{\text{pp}} + U^{\text{wp}} \right) + v_0{\hat{\bf{e}}}_{i}+ \sqrt{2D}{\bm{\eta}}_i, 
\end{equation}
\begin{figure}
	\begin{center}
		\includegraphics [width=0.75\columnwidth] {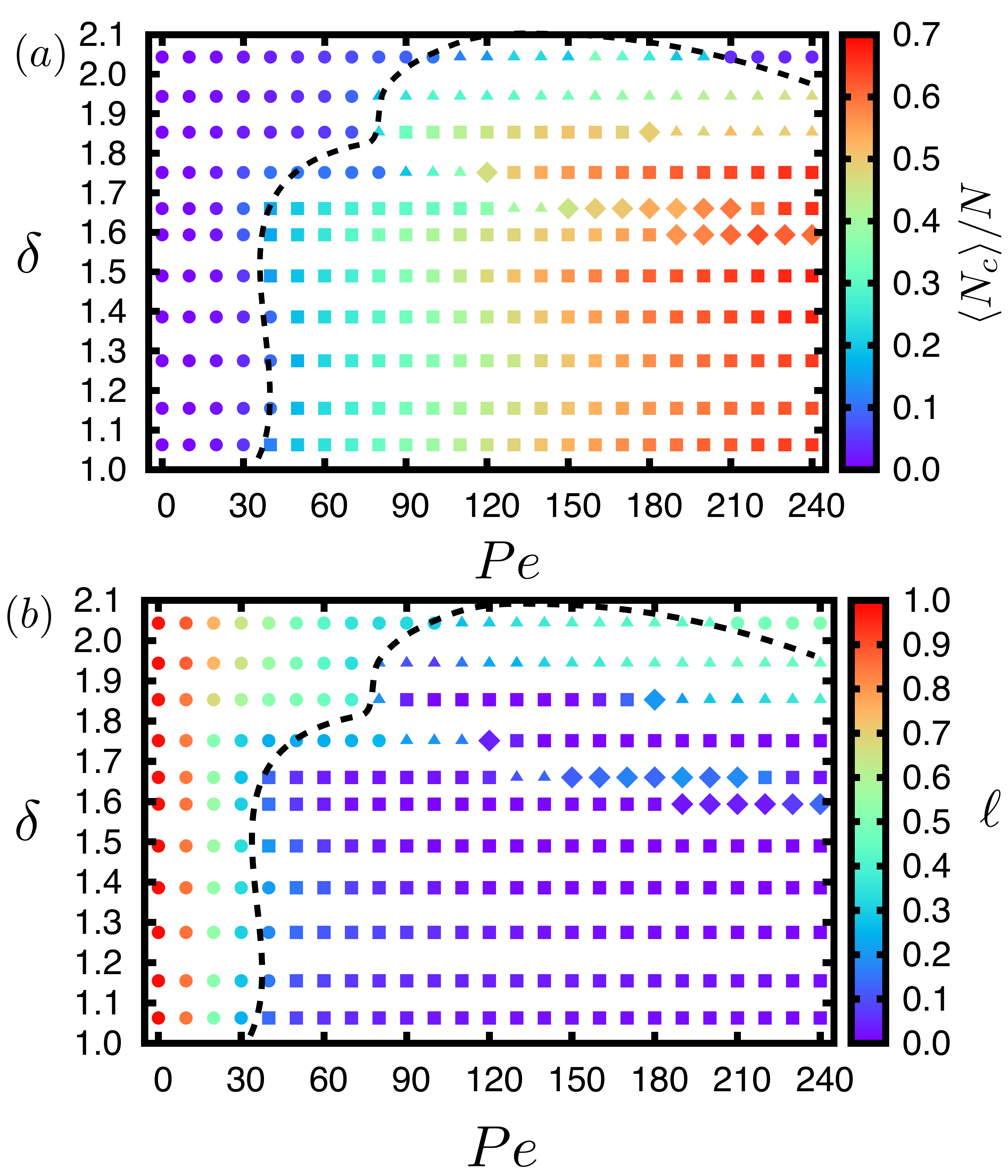}
      
		\caption{\label{fig:2} State diagram for different pore sizes $\delta$ and motility via $Pe$. Four different behaviors can be visualised here as (i) negligible clusters ($\bullet$) with $N_c \lesssim 0.1$, (ii) connected dense-phase ($\blacksquare$), (iii) bistable behaviour ($\blacklozenge$), and (iv) disconnected droplets ($\blacktriangle$). This was done for $N = 9900$ active particles for $5$ independent runs. {The color coding in (a) denotes the cluster fraction, $\langle N_c \rangle/N$(b) The same phase diagram with a colour coding denoting empty space fraction $\ell$. $\ell \simeq 0$ indicates CDP, and $\ell >0 $ indicates droplet state. Dashed curve is a guide for the eye and separates the region where wall-aggregation takes place}. 
		} 
	\end{center}
\end{figure}
$D = \mu k_BT$, is the diffusion coefficient, and $\eta$ is Gaussian white noise such that $\langle \eta(t) \rangle$ = 0 and $\langle \eta_{i\alpha}(t)\eta_{j\beta}(t^\prime)=\delta_{ij}\delta_{\alpha\beta}\delta(t-t^\prime)$. The self-propulsive speed 
$v_0$, acts along $\hat{\bf e}_i = ( \cos\theta_i, \sin\theta_i)$, where $\theta_i$  evolves as $\dot{\theta}_{i} = \sqrt{2D_{r}}\eta_{i}^{R}$ where $D_r = 3D/\sigma^2$, is the rotational diffusion coefficient. We measure the distances in unit of WCA parameter $\sigma$, time in $\tau = {\sigma^2}/{D}$, and energy in $k_BT$. {We also set the WCA potential interaction strength to unity}. We parameterize the activity by a dimensionless P{\'e}clet number $Pe = {v_0 \sigma}/{D}$ and porosity by $\delta = d/\sigma$. For all the simulations, we choose the area fraction $\phi \simeq 0.3$, in order to avoid the phase separation away from the wall~\cite{Redner2013, Fily2012}. {With this constraint, we study the system by varying the number of particles from $N = 2450$ to $249500$ ($L=80$ to $808$).} However, we conduct most of the analysis at $N=9900$ and $N=39800$. We run the simulations maximum up to time $1400 \tau$ with simulation time step  $10^{-5} \tau$. Each parameter values are examined up to 120 independent simulations. 
 When $\delta < 1$, the steric hindrance due to wall particles does not permit ABPs to penetrate through the wall, leading to a dense-phase formation of ABPs on the wall at large enough $Pe$. On the other hand, when $\delta \gg 1$, ABPs penetrate through the wall with minimal resistance, hence no particle aggregation on the wall. For $1.0 \lesssim \delta \lesssim 2.0$, the pores only permit a single ABP to pass at a time, which can cause oppositely moving particles to clog at the pores, causing a dense-phase formation.
We focus our study at this intermediate range of $\delta$.

To quantify the steady-state properties of the dense-phase, we calculate the fraction of ABPs that are part of the dense-phase, which we call the cluster fraction, $N_c$ (see \cite{SI} for details).
Subsequently, we calculate the local `height' ($h(x)$) of the dense-phase on the wall, from the $y$ coordinate of the ABP within the dense-phase at the largest $y$ separation from a given point $x$ on the wall. These measurements also provide the location of the interface between the dense-phase and the dilute phase (see \cite{SI}). When the dense-phase uniformly spreads (forms CDP) on the wall, $h \gg 1$ everywhere (Fig ~\ref{fig:1}(a)), while localized dense-phases (droplets) leave `empty' regions with $h \simeq 0$ (Fig~\ref{fig:1}(b)). Thus, we use the fraction of the empty region on the wall ($\ell$) as an identifier for different dense-phase morphologies. Considering the fluctuation in $\ell$ at transition points, we identify the dense-phase to be connected if $\ell < 0.1$. Fig~\ref{fig:1}(c-e) indicates the temporal evolution of the cluster for various parameters via $\ell$. For most parameters, the dense-phase geometry converges either to stable CDP ($\ell \simeq 0$) or to stable droplets ($\ell \gtrsim 0.3$) within a relatively short time ($t<10$). However, in a few cases, the cluster shape shows a probabilistic behavior for a finite duration such that a major fraction of the independent simulations converge to CDP while the remaining simulations show droplet formation. In many such cases, we observe late-time transitions from CDP to droplets (Fig~\ref{fig:1}(c), MOVIE\S3 \cite{SI}) or in the reverse direction (Fig~\ref{fig:1}(d), MOVIE\S4 \cite{SI}) leading to convergence to one of the two states. However, we also observed that for a narrow range of parameters, these late-time transitions become exceedingly rare, and the clusters show a robust bistable behavior (Fig~\ref{fig:1}(e)). 

In Fig.~\ref{fig:2}, we summarise the properties of ABP condensate as a function of $\delta$ and $Pe$ for $N=9900$. For $\delta \lesssim 1.5$, we only observe CDP formation at large enough $Pe$. At a higher porosity ($\delta \gtrsim 1.6$), the ABPs also form droplets. The bistable behaviour is observed in the range of $1.59 \lesssim \delta \lesssim 1.85$ at high $Pe$.  We have also analyzed the dependence of these morphological states on finite system sizes for selected parameter values corresponding to both the morphological states. We have found that both CDP and droplets persist for all the system sizes. However, the bistable region in the phase diagram shifts to different parameter values when the system size is increased. (see \cite{SI} for details).

\begin{figure}
	\begin{center}
		\includegraphics [width=\columnwidth] {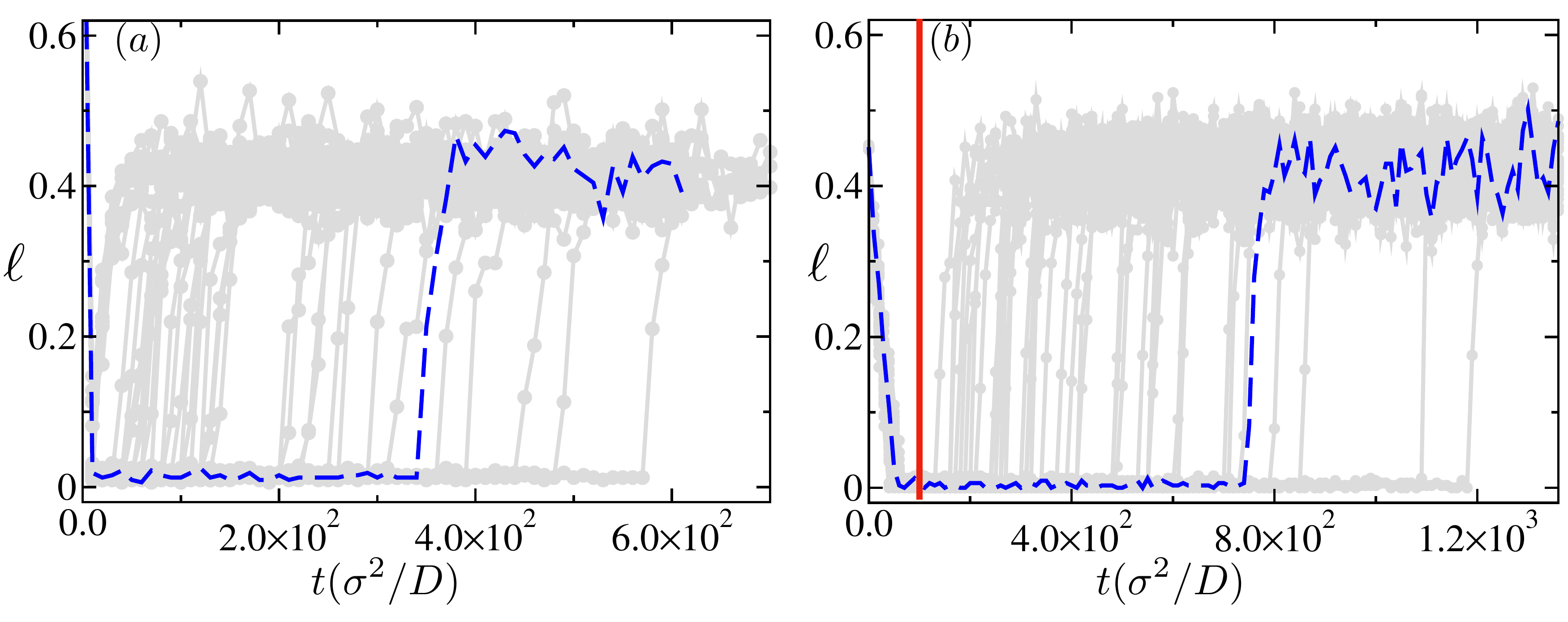}
		\caption{\label{fig:3} Vacancy fraction, $\ell$ as a function of time for (a) nucleation `seed' placed after $L/4$ at  $\delta = 1.85$ and $Pe = 240$. The seeds induced CDP, ($\ell \simeq 0$) transforms to droplets ($\ell \simeq 0.4$) (\cite{SI}-Movie\S5). (b) Time evolution of $\ell$ for a time-dependant particle motility at $\delta =1.85$. At $t=0$ the dense-phase form droplets ($\ell\simeq0.4$). The simulations converge to CDP ($\ell=0$) within $t<70$ at $Pe=120$. 
		At $t=100$ (red vertical line) the motility is escalated to $Pe = 240$, causing CDP to transform to droplets (\cite{SI}-Movie\S6). These simulations were run for $N = 39800$ for $40$ independent runs. The grey lines are the trajectories from individual simulations, and the blue line is an example of a simulation trajectory.} 
	\end{center}
\end{figure}

{\it Cluster stability and reversibility:} 
To verify that the droplet states are genuine, not a transient state which slowly evolves to a CDP, we conduct the following tests.
First, we artificially induce CDP for parameters where dense-phase form droplets, by placing nucleation `seeds' at regular intervals. These seeds are planted by blocking two adjacent pores on the wall in such a way that they close a region of length {$3 \sigma + 2 d$}. Placing the seeds on the wall reduces $d$ locally and induces local dense-phase formation. In Fig~\ref{fig:3}(a) we plot the fraction of empty region on the wall ($\ell$) for $40$ independent runs  with $\delta = 1.85 $, $Pe = 240$ with seed separation $L/4$ ($\sim 80 \sigma$). When the seeds are sufficiently close, these local clusters merge together to form CDP.  Initially, $\ell$ drops to zero within a short time ($t \lesssim 20$), indicating the formation of a seed-induced CDP in all the $40$ runs. However, at later times, all the continuous clusters spontaneously switch to droplets as $\ell$ converges to a non-zero value (Fig~\ref{fig:3}(a), Movie\S5 \cite{SI}), indicating that CDP is unstable for these parameters and droplets are indeed the preferred state.
If the droplet states were transient and CDP is the globally preferred state, the induced CDP would have remained stable in these simulations. However, the late-time transitions from CDP to droplets in these simulations assert the robustness of the droplet states. Repeated tests at smaller seed separations (L/5) also show similar behaviour.  To further verify the robustness of both the morphological state, we also study the coarsening behaviour of the aggregates of these states by calculating the size of the largest cluster as a function of time~\cite{Redner2013} (\cite{SI}). In all these cases, we do not observe any increase in the aggregate size beyond $t\simeq 10$, where a single large cluster is formed, again confirming that cluster shapes are not transient states. Next, we study the time-evolution of aggregates by varying the particle motility with time (Fig~\ref{fig:3}(b)). We choose a droplet as an initial configuration and conduct the simulation for $Pe$ and $\delta$, where CDP is apparently stable. The time-evolution of $\ell$ (Fig~\ref{fig:3}(b) and Movie\S6 \cite{SI}) in such a simulation for $\delta =1.85$ at $Pe=120$ indicates that the initial droplets transform into CDP, as $\ell$ drops to zero within a relatively short time, $t<70$. Afterwards, at $t=100$ (vertical line in Fig~\ref{fig:3}(b)), motility is increased to a higher value ($Pe=240$). We find that the CDP destabilizes at this motility and transforms into droplets. These tests confirm the motility dependence of both the morphological states. Also, these results show that time-dependent motility dynamically alters the dense-phase morphology on the wall.

{\it Wall-liquid interfacial tension:} The ABP aggregation on the wall leads to the formation of at least two different types of interfaces, one at the boundary between dense(liquid) and the dilute(vapor) phases and the second one separating the wall and the liquid phase. It has been shown previously that the presence of interfaces changes the mechanical stress distribution in active-materials~\cite{Bialke2015,paliwal2017non}. To understand the modifications in stress distributions due to walls, we calculate the components of the pressure tensor as a function of the distance from the wall. The self propulsion of ABPs contributes to the swim pressure, 
\begin{equation}
  p_{\alpha \beta}^{(s)}(y) = \dfrac{1}{2L\delta_y}\langle \sum_{i \in \delta_y} j_{i\alpha}v_{i\beta}\rangle, 
\end{equation}
where $\langle.\rangle$ indicates time-average within rectangular bins of width $\delta_y$  aligned parallel to the wall and to the interfaces, $\alpha,\beta$ are the Cartesian components. $\bm{ j}_i$ denotes the active impulse given by $v_0{\hat{\bf{e}}}_{i}/\mu D_r$ and $\bm{v}_i$ the velocity of the $i$-th particle \cite{Fily_2017, das2019local}.
 The stress caused by inter-particle interactions $F_{ij}$, separated by a distance $r_{ij}$ is given as,
\begin{equation}
   p_{\alpha \beta}^{(I)}(y) = \dfrac{1}{2L\delta_y} \langle \sum_{(i/j) \in \delta_y} F_{ij\alpha}r_{ij\beta}\rangle,
\end{equation}
Extending the equilibrium studies~\cite{nijmeijer1990wetting}, the stress due to wall interaction $F_{iw}$ is,
 \begin{equation}
    p_{\alpha \beta}^{(w)}(y) = \dfrac{1}{L\delta_y} \langle \sum_i F_{iw\alpha}r_{iw\beta}\rangle,
 \end{equation}
{where $r_{iw}$ is the distance of the $i$-th active particle with the $w$-th wall particle.}
 When the dense-phase forms CDP, the non-vanishing, diagonal components $p_{xx}$ and $p_{yy}$ become the tangential ($p_T$) and the normal ($p_N$) components to the interfaces. In the limit of large $Pe$, we neglect the thermal stress contributions. 

\begin{figure}
    \centering
    \includegraphics [width=\columnwidth]{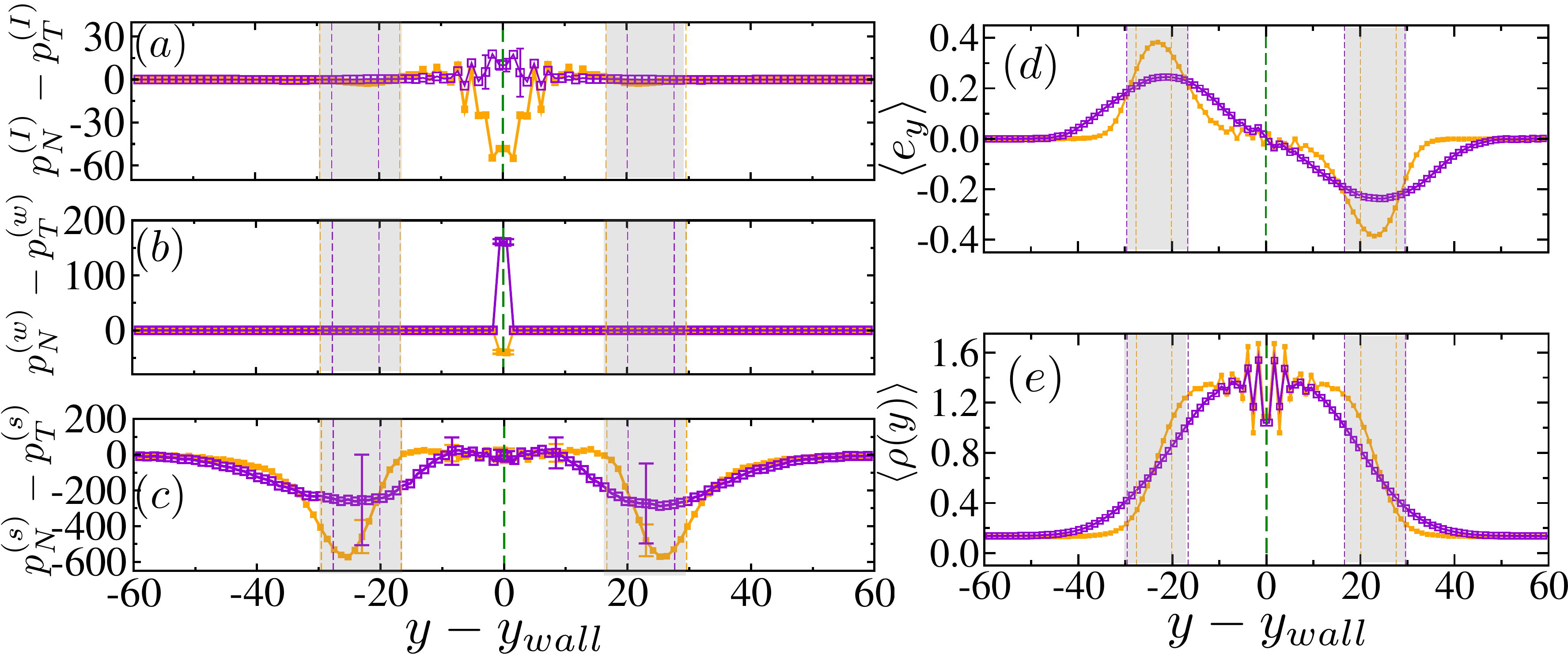}
    \caption{Pressure difference $(p_N - p_T)$ arising from (a) interaction pressure, (b) wall interaction and (c) swim contribution as a function of distance from wall for $\delta = 1.59, Pe = 240$ (orange curve) and $\delta = 1.85, Pe = 240$ (violet curve). The swim pressure difference peaks near the liquid-vapor interface, while the interaction and wall pressure difference contributes in the dense-phase. (d) Distributions of particle orientation normal to the wall, and (e) density distribution, as a function of normal distance from the wall for $\delta =1.59$ (orange) and $\delta = 1.85$ (violet) at $Pe=240$. Simulations are performed for $N = 39800$. The shaded region indicate the location of liquid-vapour interface.}
    \label{fig:4}
\end{figure}

To understand the effect of wall properties in the dense-phase morphology, we compare the pressure distribution for the same $Pe$, but at different $\delta$, where we observe contrasting aggregate morphology. Thus, we closely analyse systems at $Pe = 240$, which provides a stable CDP at $\delta = 1.59$, but form droplets at $\delta = 1.85$.  Also, in the latter case, the dense-phase remains in CDP for a significant amount of time (cf. Fig.~\ref{fig:1}(c)), before they break-up into droplets. The geometric similarity between the stable CDP ($\delta =1.59$) and the `unstable' CDP $\delta= 1.85$, allows us to directly compare the distribution of $p_{N/T}$ in these cases. We choose those configurations which remain as CDP for this comparison. {For such a geometry, one obtains the total surface tension, including contributions from all the four interfaces by extending the approach used in equilibrium wetting phenomena~\cite{Kirkwood1949,nijmeijer1990wetting},
\begin{equation}
  \gamma_{wl}+\gamma_{lv} = {1\over 2}\int_{-L/2}^{L/2} \{(p^{(I)}_{N}-p^{(I)}_{T})+(p^{(w)}_{N}-p^{(w)}_{T})+(p^{(s)}_N-p^{(s)}_T)\}  dy.
    \label{eq:5_n}   
\end{equation}
}
Thus we calculate the difference, ($p_N - p_T$) for the three pressure contributions for $\delta = 1.59$ and $\delta =1.85$ as a function of $y$ as given in Fig~\ref{fig:4}(a)-(c). Interestingly, we find significant difference ($p_N^{(I)}-p_T^{(I)}$) inside the dense region and away from the wall (Fig.~\ref{fig:4}(a)), which eventually vanishes near the liquid-vapour interface. This behaviour is in contrast with the observations inside dense phase without the walls, where ($p_N^{(I)}-p_T^{(I)}$) vanishes in the bulk~\cite{Bialke2015}(see \cite{SI}). This result indicates that, even though the wall force acts only on the first contact layer of ABPs, it alters the stress distribution over multiple layers of particles. Thus, the dense region doesn't attain bulk properties in these simulations.  Also, for $\delta=1.59$,  $(p_N^I-p_T^I) <0$ in the dense-phase while it is weakly positive for $\delta =1.85$ (Fig~\ref{fig:4}(a)). The contributions from direct wall interaction is non-zero only within a distance $\lesssim 2^{1/6}\sigma$ from the wall (Fig.~\ref{fig:4}(b)). However, the swim pressure difference, ($p_N^{(s)} - p_T^{(s)}$) is not significant in the dense-phase whereas it shows a large (negative) value near the liquid-vapour interface as observed previously (Fig~\ref{fig:4}(c)). Using these measurements, we {calculate} the wall-liquid surface tension,  
\begin{equation}
    \gamma_{wl} = {1\over 2}\int_{-L/2}^{L/2} \{(p^{(I)}_{N}-p^{(I)}_{T})+(p^{(w)}_{N}-p^{(w)}_{T})\}  dy.
    \label{eq:5}
\end{equation}
 We ignore the swim pressure contribution $(p^{(s)}_N - p^{(s)}_T)$ Eq,~\ref{eq:5}, {as it is negligible near the wall-liquid interface.} Also, the tangential component of $p^{(w)}$ does not vanish since the particles can penetrate the wall. {Similarly, we also calculate the liquid-vapour interfacial tension $\gamma_{lv} = {1\over 2} \int_{-L/2}^{L/2} (p^{(s)}_N - p^{(s)}_T) dy$, considering that $(p^{(I/w)}_N - p^{(I/w)}_T)$ has negligible contribution at the liquid-vapor interface.}
  Our calculations indicate that when $\delta = 1.85$, the wall acts as a high-energy surface ($\gamma_{wl} \simeq 237 $) and $\delta =1.59$ as a low-energy surface ($\gamma_{wl} \simeq -208$). We also observe a steady increase in $\gamma_{wl}$ with increase in $Pe$ at $\delta = 1.85$ (\cite{SI}-Table\S 2) as the dense-phase shows increasing tendency to form droplets.  Subsequently, we define a spreading parameter $S = - (\gamma_{wl}+\gamma_{lv})$ following the approach in equilibrium surface-wetting studies~\cite{deGennes2004}. Here we ignore $\gamma_{wv}$ as its values are not significant. We obtain $S(\delta = 1.59) > S(\delta = 1.85)$ for $Pe = 240$.  However, we couldn't observe this trend in $S$ when we repeat analysis by varying $Pe$ from $120$-$240$ at $\delta = 1.85$, where the preference to form droplets increase with $Pe$. We note that the value of S is influenced by $\gamma_{lv}$, which is a large negative quantity. Also, unlike the interaction pressure, $p^{(s)}$ is more difficult to estimate locally, partly due to its sensitivity to the finite system size. 

The observed difference in interaction pressure components is due to the modification of inter-particle interaction in the dense-phase in the presence of wall, as we do not observe such a difference in its absence~\cite{Bialke2015}(\cite{SI}). Since the ABP interactions are linked to the local particle orientation and the density, we compare the orientation $\langle {\hat{\bf{e}}}_{y} \rangle (y)$  (Fig.~\ref{fig:4}(d)), and the  density $\rho(y)$ distributions (Fig.~\ref{fig:4}(e)), as a function of distance from the wall $y$, for $\delta =1.59$ and $\delta =1.85$ at $Pe=240$. Evidently, an increase in wall porosity increases the width and decreases the peak value of the $\rho(y)$. {Also, we note that the $\rho(y)$ does not reach a plateau (especially for $\delta = 1.85$) as seen for a bulk phase.} We observe a similar trend in $\langle {\hat{\bf{e}}}_{y} \rangle$ (Fig.~\ref{fig:4}(d)). The orientation distribution shows a non-vanishing particle orientation inside the dense region, while such an orientational preference is localized in the liquid-vapour interface in the absence of walls (\cite{SI}). This indicates a significant influence of the wall in the dense region. Also, the properties of the dense region is different from the bulk behaviour, since the thickness of the dense layer on the wall is smaller than the effective range of influence of the wall. Since the active force along the particle orientation is balanced by the interaction force, a non-vanishing mean orientation influences the stress distribution within the aggregates, hence the change in $(p_N^{(I)} - p_T^{(I)})$. Also, a broader density distribution indicates enhanced structural disorder, triggered by particles crossing the wall, which effectively changes the interaction energy inside the dense-phase. 

In summary, we demonstrate that a morphological transition in the dense-phase on porous walls by ABPs can be achieved either by changing the wall porosity or the particle motility. This transition is reversible, and the wall spreading of the dense-phase can be tuned by altering these two quantities. Interestingly, the droplet formation on the walls coincides with an increase in effective wall-liquid interfacial tension. Higher wall porosity influences the physical properties of the dense-phase by altering the local density and the polarity, which leads to a change in interaction pressure in the dense-phase. These observations imply that the underlying reason for morphological transformation is more general and can be observed in a variety of ways, {even for impenetrable walls}. A general understanding of this phenomenon requires the development of theoretical models for active-morphological transitions. Understanding such morphological transitions can be useful for many practical applications, such as the design of surfaces that prevent microbial spreading, especially in the context of recent studies~\cite{Liu2019} revealing the role of MIPS in the aggregation of {\it Myxococcus xanthus}.

\section{Acknowledgements}
We thank Gerhard Gompper, Roland Winkler, Sriram Ramaswamy, and Julien Tailleur for discussions. RC thanks  Arvind Gopinath and Mandar Inamdar for going through the manuscript. SD thanks Debasmit Sarkar for technical help. RC acknowledges the financial support from SERB, India via the grants SB/S2/RJN-051/2015 and SERB/F/10192/2017-2018. We acknowledge SpaceTime2 HPC facility at IIT Bombay.

\bibliography{rsc}
\bibliographystyle{rsc} 

\end{document}